\documentclass{article}

\usepackage{amssymb,amsfonts,amsmath,stmaryrd}
\usepackage{cite,enumerate,float,indentfirst}
\usepackage{color}

\def\be{\begin{eqnarray}}
\def\ee{\end{eqnarray}}
\def\nn{\nonumber}

\def\p{\partial}

\def\Tr{{\rm Tr}\,}

\definecolor{red}{rgb}{1,0,0}
\definecolor{orange}{rgb}{1,0.5,0}
\definecolor{violet}{rgb}{0.7,0,1}

\hoffset= - 1.0in         % switch off for draft style
\begin{document}

\hfill ITEP/TH-01/19

\hfill IITP/TH-01/19

\bigskip

\centerline{\Large{
On $W$-representations of $\beta$- and $q,t$-deformed matrix models
}}

\bigskip

\centerline{{\bf A.Morozov}}

\bigskip

\centerline{\it ITEP \& IITP, Moscow, Russia}

\bigskip

\centerline{ABSTRACT}

\bigskip

{\footnotesize
$W$-representation realizes partition functions by an action of a cut-and-join operator
on the vacuum state with a zero-mode background.
We provide explicit formulas of this kind for $\beta$- and $q,t$-deformations
of the simplest rectangular complex matrix model.
In the latter case,
instead of the complicated definition in terms of multiple Jackson integrals,
we define partition functions as the weight-two series, made from Macdonald polynomials,
which are evaluated at different loci in the space of time variables.
Resulting expression for the $\hat W$ operator
appears related to the problem of simple Hurwitz numbers
(contributing are also the Young diagrams with all but one lines of length two and one).
This problem is known to exhibit nice integrability properties.
Still the answer for $\hat W$ can seem unexpectedly sophisticated
and calls for improvements.
Since matrix models lie at the very basis of all gauge- and string-theory constructions,
our exercise provides a good illustration of the jump in complexity
between $\beta$- and $q,t$-deformations --
which is not always seen at the accidently simple level of
Calogero-Ruijsenaars Hamiltonians (where both deformations are equally straightforward).
This complexity is, however, quite familiar in the theories of network models,
topological vertices and knots.
}

\bigskip

\bigskip

\section{Introduction}

Eigenvalue matrix models \cite{MAMO}
play an increasingly important role in modern theory.
Originally they appear as convenient representatives of universality classes
of important statistical distributions (random matrix ensembles) \cite{Dyson}
and as exactly solvable examples of quantum field and string theory models \cite{Migdal}.
Today they are also considered as exhaustively describing the non-perturbative
sector, where they appear through either
the ADHM description of instanton calculus \cite{ADHM},
or the free-field formalism in conformal theory \cite{CFT}
or as the outcome of localization methods \cite{loc}.
A variety of important applications stimulates the study of matrix models
{\it per se}, which revealed a lot of important structures,
widely used in modern physics, and points to the hidden matrix-model
structures far beyond their traditional areal.
At early stages the central properties seemed to be integrability
and exhaustive set of Ward identities (Virasoro constrains),
reviewed in \cite{UFN3} and later promoted to a powerful method
of the AMM/EO topological recursion \cite{AMM/EO}.
Today they are considered as consequences of a more fundamental
"superintegrability" property $\boxed{\langle{\rm character}\rangle=character}\ $ \cite{MMchar},
which survives also in tensor models and, perhaps, even knot theory.
One more conceptually important description is in terms of evolution
in the moduli space of coupling constants -- known in matrix-model context
as W-representations \cite{MShWreps}.

The goal of this paper is to make a step towards unification of these
two approaches.
We work out a $W$-representation of $q,t$-models, which are the first example,
where the superintegability-based definition from \cite{MPSh} is indisputably simpler than the
conventional one \cite{HS5d} through a matrix-like Jackson integral.
In this case the latter would be a network-model lifting \cite{networkZ,network}
of Dotsenko-Fateev matrix model \cite{CFT},
which is a free-field description of intertwiner (topological vertex) convolutions
in DIM algebra \cite{DIM}, where integrals over the matrix eigenvalues
(which are the arguments of the screening integrands)
are actually substituted by sums over Young
diagrams and, further, over plane partitions.
The former definition, at least for the simplest representations of DIM,
is just a relatively easy sum  over Macdonald functions
\cite{MPSh}, and we will show that its W-representation is, as expected,
a direct generalization of the formula for traditional matrix models.
It is governed by a
relative of Calogero-Ruijsenaars Hamiltonian \cite{CaRu}
(which is the $\beta$ and $q,t$-deformation of the simplest of cut-and-join operators
from \cite{MMN1}).
However, for generic $q$ and $t$ this "relative" appear to be far more sophisticated
and is not just a naive "neighbor  harmonic" of the same operator
(like it remains to be at the level of  $\beta$-deformation).
In this paper we provide just a raw formula, which deserves further
rewriting and better understanding.

\section{The definition of Gaussian $q,t$-model}

For ordinary Gaussian matrix models one can explicitly calculate
any particular correlator (and various generating functions of those)
and {\it observe} a spectacular fact, that averaging
preserves the {\it shape} of some functions, i.e.
a function of original ("microscopical") variables  (fields)
after ("functional") integration becomes {\it the same} function
of the final ("macroscopical") variables.
Moreover, these eigenfunctions of "renormalization group evolution"
in matrix models are very simple: just the Schur functions,
i.e. the characters of the underlying symmetry.
Exact statement in the case of the two simplest,
Gaussian Hermitian and complex matrix \cite{compmat} models, are \cite{MMmamo}:
\be
Z_N\{p\} = c_N\int dM e^{-\frac{1}{2}\Tr M^2+\sum_k \frac{ p_k\Tr M^k}{k}}
= \sum_{R} \frac{{\rm Schur}_R\{p_k=\delta_{k,2}\}\cdot {\rm Schur}_R\{p_k=N\}}
{{\rm Schur}_R\{p_k=\delta_{k,1}\}} \cdot {\rm Schur}_R\{p\}
\nn \\
Z_{N_1\times N_2}\{p\} =
c\!\int d^2M e^{-\frac{1}{2}\Tr (MM^\dagger)+\sum_k \frac{ p_k\Tr (M M^\dagger)^k}{k}}
= \sum_{R} \frac{{\rm Schur}_R\{p_k=N_1\}\cdot {\rm Schur}_R\{p_k=N_2\}}
{{\rm Schur}_R\{p_k=\delta_{k,1}\}} \cdot {\rm Schur}_R\{p\}
\label{sumsR}
\ee
This fact does not follow directly neither from integrability nor
from the Virasoro constraints -- instead they both are its corollaries.
Thus it can be considered as the long-hunted for formulation of
the {\it superintegrability} property, of which the above two are
the two {\it complementary} implications.

Perhaps, even more interesting is the simplicity of the formulas (\ref{sumsR}),
which calls for all kinds of generalizations --
and one of the most interesting for today's applications is
to $q,t$-models, which are now intensively studied my far more
complicated techniques.
Instead, according to \cite{MPSh}, partition functions of the Gaussian  $q,t$-models
can be just {\it postulated} to be
%given by
\be
Z\{p\} = \sum_R \frac{\Big<{\cal M}_R\{P\}\Big>}{||{\cal M}_R||^2}\cdot {\cal M}_R\{p\}
= \left<\exp\left(\sum_k  \frac{t^{k}-t^{-k}}{q^{k}-q^{-k}}\cdot p_k P_k\right)\right>
\label{charavMC}
\ee
where ${\cal M}_R\{p\}$ are Macdonald polynomials \cite{Macdonald}, and
average over the fields $P_k$ is defined through
\be
\Big<{\cal M}_R\{P\}\Big>_{N_1\times N_2}
= \frac{{\cal M}_R\{\pi^{(N_1)}\}{\cal M}_R\{\pi^{(N_2)}\}}{{\cal M}_R\{\delta^*_{k,1}\}}
\label{charavMH}
\ee
and
\be
\Big<{\cal M}_R\{P\}\Big>_{N}
= \frac{{\cal M}_R\{\delta^*_{k,2}\}}{{\cal M}_R\{\delta^*_{k,1}\}}\cdot {\cal
M}_R\{\pi^{(N)}\}
\ee
respectively for the complex and Hermitian models.
Here the topological locus and the important loci in the space of time variables $\{p_k\}$ are
\be
\pi^{(N)}_k = \frac{t^{Nk}-t^{-Nk}}{t^k-t^{-k}}
\ \ \stackrel{t=q^\beta\longrightarrow 1}\longrightarrow \ \ N
\ee
and
\be
\delta^*_{k,n} =
n\cdot \frac{(q-q^{-1})^{k/n}}{t^k-t^{-k}}\cdot \delta_{k| n}
\ \ \stackrel{t=q^\beta\longrightarrow 1}\longrightarrow \ \
\frac{\delta_{k,n}}{\beta}
\label{delta*}
\ee
where $\delta_{k| n}=1$ when $k$ is divisible by $n$ and zero otherwise.
We showed also what happens in the limit when both $t$ and $q$ tend to unity,
but $t=q^\beta$ -- this is what is usually called $\beta$-deformation
\cite{betadefo}
(in Nekrasov calculus $\beta = -\epsilon_2/\epsilon_1$).
Alternatively $q,t$-model can be defined as a free-field correlator
\cite{HS5d} in DIM-based
network models \cite{networkZ}, see also \cite{network} --
but exact comparison remains to be done.
There the time-variables $P_k$ are actually the fields,
generalizing the standard matrix model expressions
$P_k=\Tr (XX^\dagger)^k$ and $P_k = \Tr X^k$
in complex and Hermitian cases.
Note that Gaussian model is different from the logarithmic (Dotsenko-Fateev) one, 
directly relevant for the AGT relation \cite{AGT}, where the analogue of the basic
relations (\ref{charavMH}) and (\ref{charavMC}) involves
{\it generalized} Macdonald functions at the l.h.s. (with the deformation parameter
 $\epsilon=\epsilon_1+\epsilon_2$, measuring the deviation from $\beta=1$) 
and Nekrasov functions at the r.h.s., see \cite{HS5d,MSmi,MaMZ,Zgenmac} 
for the step-by-step solution
of that model and the resulting exhaustive proof of the AGT relation
on the lines of \cite{Ito} and \cite{MMSha}.

In this paper we look for still another definition/reformulation of the Gaussian
$q,t$-model:
in terms of the $W$-representations \cite{MShWreps}, i.e.
represent the partition functions (\ref{sumsR}) as
\be
Z\{p\} = e^{\hat W\{p\}}\cdot 1
\ee
with operator $\hat W$ depending on $p$-variables and $p$-derivatives $\p/\p p_k$.

\section{$\beta$-deformation of Gaussian Hermitian and complex models ($t=q^\beta$) }

To avoid writing extra formulas, we present here the $W$-representation in
the $\beta$-deformed case:
it is easy  to return to the original
result of \cite{MShWreps} and \cite{compmod} by putting $\beta=1$.
and substituting Jack polynomials  by the ordinary Schur functions.
For examples of $W$-representations in other conventional
models see \cite{Alwrep,AMMN} and \cite{MMchar}
(it deserves mentioning that the celebrated Rosso-Jones formula for torus knots \cite{RJ,RJnontor}
is also an example).

The $\beta$-deformed Gaussian Hermitian model (also known as Gaussian $\beta$-ensemble)
was described in detail in \cite{MMPSh}.
It can be defined as an eigenvalue  integral
\be
\tilde Z_\beta^{(N)}\{p\} \sim \int   \prod_{i<j}^N (x_i-x_j)^{2\beta}
\prod_{i=1}^N \exp\left(\sum_k \frac{p_kx_i^k}{k}\right) e^{-x_i^2/2} dx_i
\ee
The averages, calculated in \cite{MMPSh},
are reproduced by the straightforward $\beta$-deformation of the
standard $W$-representation of  \cite{MShWreps}:\footnote
{I am indebted to Tomas Prochazka and Piotr Sulkowski for raising the question
about this old unpublished complement to \cite{MShWreps} and \cite{MMPSh}
-- and to Shamil Shakirov for confirming the formula.
}
\be
\!\!\!\!\!\!\!\!\!\!\!\!\!\!\!
\tilde Z_\beta^{(N)}\{p\} = e^{-Nt_0} \cdot \exp\left\{
\frac{1}{2}\sum_{a,b=0}\left( abt_at_b\frac{\p}{\p t_{a+b-2}}
+\beta\cdot(a+b+2)t_{a+b+2}\frac{\p^2}{\p t_a\p t_b}\right)
+\frac{1-\beta}{2}\sum_{a=0}(a+1)(a+2)t_{a+2}\frac{\p}{\p t_a}
\right\} \cdot e^{Nt_0}
= \nn
\ee
\vspace{-0.4cm}
\be
\!\!\!\!\!\!\!\!
= \exp\left\{ \frac{(\beta N  -\beta+1  )Np_2 + Np_1^2 }{2}
%+ \phantom{\left(\sum_{a=0}a(a+1)p_{a+2}\frac{\p}{\p p_a}\right)}
%\right.\nn \\ \left.
+\frac{1}{2}\sum_{a,b=1}\left(
(a+b-2)p_ap_b\frac{\p}{\p p_{a+b-2}}
+\beta\cdot abp_{a+b+2}\frac{\p^2}{\p p_a\p p_b}  \right)
+ \right.\nn \\ \left.
+ \beta N \sum_{a=1} ap_{a+2}\frac{\p}{\p p_a}
+\frac{1-\beta}{2}\left(\sum_{a=1}a(a+1)p_{a+2}\frac{\p}{\p p_a}\right)^{\phantom{5}}\!\!\!
\right\}\cdot 1  = e^{\hat W_\beta^{(N)}}\cdot 1
\ee
where $W$-operator in the exponent is the $-2$-th harmonic of the
the Calogero-Ruijsenaars Hamiltonian \cite{CaRu},
e.g. of (23) in \cite{MSmi}, and we presented expressions in terms of both
popular choices of time-variables: $t_k$ and $p_k=kt_k$.

The character expansion of \cite{MMmamo} in this case is in terms of Jack polynomials
${\rm Jack}_R\{p\} = \left.{\cal M}_R\{p\}\right|_{t=q^\beta\longrightarrow 1}$:
\be
\tilde Z_\beta^{(N)}\{p\} = \sum_R \ \underbrace{\beta^{|R|/2}\cdot
\frac{{\rm Jack}_R\{\delta_{k,2}\}\cdot{\rm Jack}_R\{p_k=N\}}{{\rm Jack}_R\{\delta_{k,1}\}}}_{
\Big<{\rm Jack}_R[X]\Big>}
\cdot \frac{{\rm Jack}_R\{\frac{p_k}{\beta}\}}{||{\rm Jack}_R||^2}
\label{tildeZbetathroughJacks}
\ee
and contributing, as usual in Gaussian Hermitian models, are only Young diagrams
$R$ of even size $|R|$.
A typical example of the entries in this expression:
\be
{\rm Jack}_{[3,2,1]}\{p_k=N\} =
\frac{(\beta N+2)(\beta N+1) N(N-1)(N-2) (\beta N -\beta+1)}{(2\beta+1)^2(3\beta+2)}
\nn \\
{\rm Jack}_{[3,2,1]}\{p_k=\delta_{k,1}\} =\frac{\beta^3}{(2\beta+1)^2(3\beta+2)},
\ \ \ \ \ \ \
{\rm Jack}_{[3,2,1]}\{p_k=\delta_{k,2}\} = \frac{\beta(\beta-1)}{(2\beta+1)^2(3\beta+2)}
\ee
In particular, one can see that the vanishing Gaussian average
$\Big<{\rm Schur}_{[3,2,1]}[X]\Big>=0$ becomes non-vanishing after a $\beta$-deformation
to $\ \Big<{\rm Jack}_{[3,2,1]}[X]\Big>_\beta\sim (\beta-1)$.

We wrote eq.(\ref{tildeZbetathroughJacks}) in conventional notation of \cite{MMPSh}.
To make it  consistent with the general expression in the present text one should
rescale $p_k\longrightarrow \beta p_k$, to get an expansion in ${\rm Jack}\{p\}$:
\be
Z_\beta^{(N)}\{p\} := \tilde Z_\beta^{(N)}\{\beta p\}
= \sum_R  \beta^{|R|/2}\cdot
\frac{{\rm Jack}_R\{\delta_{k,2}\}\cdot{\rm Jack}_R\{p_k=N\}}{{\rm Jack}_R\{\delta_{k,1}\}}
\cdot \frac{{\rm Jack}_R\{p\}}{||{\rm Jack}_R||^2}
\ =\ e^{\hat W_\beta^{(N )}}\cdot 1 \ = \nn \\
= \exp\left\{ \frac{(\beta N  -\beta+1  )\beta Np_2 + \beta^2 Np_1^2 }{2}
+\frac{1}{2}\sum_{a,b=1}\left(
\beta\cdot (a+b-2)p_ap_b\frac{\p}{\p p_{a+b-2}}
+  abp_{a+b+2}\frac{\p^2}{\p p_a\p p_b}  \right)
+ \right.\nn \\ \left.
+ \beta N \sum_{a=1} ap_{a+2}\frac{\p}{\p p_a}
+\frac{1-\beta}{2}\left(\sum_{a=1}a(a+1)p_{a+2}\frac{\p}{\p p_a}\right) \right\}\cdot 1
\label{betaC}
\ee
Likewise, for the $\beta$-deformation of the complex model we have
\be
\sum_R \ \beta^{|R|}\cdot
\frac{{\rm Jack}_R\{p_k=N_1\}\cdot{\rm Jack}_R\{p_k=N_2\}}{{\rm Jack}_R\{\delta_{k,1}\}}
\cdot \frac{{\rm Jack}_R\{p\}}{||{\rm Jack}_R||^2} \
\ =\ e^{\hat W_\beta^{(N_1\times N_2)}}\cdot 1\  = \nn \\
  = \exp\left\{ \beta^2 N_1N_2p_1
+ \sum_{a,b=1}\left(
\beta\cdot (a+b-1)p_ap_b\frac{\p}{\p p_{a+b-1}}
+  abp_{a+b+1}\frac{\p^2}{\p p_a\p p_b}  \right)
+ \right.\nn \\ \left.
+ \beta (N_1+N_2) \sum_{a=1} ap_{a+1}\frac{\p}{\p p_a}
+(1-\beta)\left(\sum_{a=1}a^2 p_{a+1}\frac{\p}{\p p_a}\right) \right\}\cdot 1
\label{WJC}
\ee
These are the two formulas which we want to lift to the $q,t$
%(and, probably, one day, to Kerov)
models.
Note that the weights $\beta^{|R|/2}$ and $\beta^{|R|}$ at the l.h.s. of these formulas
automatically appear in the limit
$\ t=q^\beta\rightarrow 1\ $
from the factor $\frac{1}{\beta}$
in (\ref{delta*}).

\section{On ambiguity of $W$-representation
\label{ambigbeta}}

In fact, $W$-representation is ambiguous in the following sense.
If we only demand  that the l.h.s. of, say, (\ref{WJC}) at $\beta=1$, is reproduced
term-by-term by the action of some evolution operator on unity,
then the operator at the r.h.s. of (\ref{WJC}) can be modified by adding to the exponent
of any linear combination of infinitely-many new operators.
The simplest of them is
\be
\sigma_3\{p\}\cdot \left(2(N_1N_2+1)\cdot\frac{\p}{\p p_2} - (N_1+N_2)\cdot\frac{\p^2}{\p
p_1^2}
- \frac{4(N_1^2-1)(N_2^2-1)}{(N_1N_2+4)(N_1+N_2)^2+ 2(N_1N_2+1)}
\cdot p_2\frac{\p^2}{\p p_2^2}+ \ldots\right)
\label{ambWJC3}
\ee
with a coefficient, which arbitrarily depends on $p_k$ --
if one prefers to preserve grading, it should be
restricted to
$\sigma_3\{p\} = \sigma_{_{[3]}} p_3 + \sigma_{_{[2,1]}}p_2p_1+\sigma_{_{[1,1,1]}}p_1^3$.
At the level $4$ there are two new operators with arbitrary coefficients:
\be
\sigma_{4a}\{p\}\cdot\left(3(N_1N_2+1)(N_1N_2+2)\cdot\frac{\p}{\p p_3}
- (N_1^2+3N_1N_2+N_2^2+1)\cdot\frac{\p^3}{\p p_1^3}+\ldots\right)
+ \nn \\
+\sigma_{4b}\{p\}\cdot\left(2(N_1N_2+1) \cdot\frac{\p^2}{\p p_2\p p_1}
- (N_1+N_2)\cdot\frac{\p^3}{\p p_1^3}+\ldots\right)
\label{ambWJC4}
\ee
and so on.
Note that the second operator in (\ref{ambWJC4}) resembles a $p_1$-derivative of
(\ref{ambWJC3}).

One can fix this freedom by restricting the $N$-dependence of $\hat W$ to contain just three
structures: $N_1N_2$, $N_1+N_2$ and $1$ --
this picks up the nice formula (\ref{WJC})
and forbids all the ugly ambiguities.

\section{Towards $W$-representation for $q,t$-models}

While $\beta$-deformation of $W$-representation is very simple,
this is not quite the case for the generic $q$ and $t$
and we construct it in five steps.

 First will be specification of $N$-dependence, see eq.(\ref{Ndep}) below.

 Second is finding the first contributions to the $N$-independent
part $\hat W$ of the evolution operator, see (\ref{firstterms}).

 Third -- the study of its ambiguities, see sec.\ref{ambigqt}.

 Forth -- understanding the peculiar tri-linear structure (\ref{trilin})
of $\hat W$ and

  Fifth -- providing a generating function (\ref{genfun}) for its generic term.

The most surprising in this story is that one expects the relevant ${\cal W}$
operator to be a close relative of Calogero-Ruijsenaars Hamiltonian for generic $q$ and $t$,
\be
\hat{ H} =
\oint\frac{dz}{z}\exp\left(\sum_{k=1} \frac{(1-t^{-2k})z^kp_k}{k}\right)
\exp\left(\sum_{k=1}\frac{q^{2k}-1}{z^k}\frac{\p}{\p p_k}\right)
= \nn \\
= \sum_{m=0} t^{-2m}\cdot {\rm Schur}_{[m]}\Big\{(t^{2k}-1)p_k\Big\}
\cdot{\rm Schur}_{[m]}\left\{(q^{2k}-1)k\frac{\p}{\p p_k}\right\}
\label{HamCR}
\ee
-- the simplest in the operator family, of which Macdonald polynomials are the eigenfunctions.
However, to get the needed ${\cal W}$ operator, which provides
the $W$-representation of a $q,t$-model, it is not enough just to change
the power of $z$ in $\oint\frac{dz}{z}$.
The answer is substantially more involved --
what, of course, is {\it not} really a surprise for experts in $q,t$-deformations.
The structures, revealed in this basic exercise, can now be expected
to emerge in more sophisticated applications.

To avoid repeating the lengthy story twice we concentrate in this paper
on the $q,t$-deformation of the complex model, and our target is the operator
${\cal W}$ in
\be
{\cal Z}_{N_1\times N_2} = \sum_{m=0} {\cal Z}_m \equiv
\sum_R
\frac{{\cal M}_R\{\pi^{(N_1)}\}\cdot{\cal M}_R\{\pi^{(N_2)}\}}{{\cal M}_R\{\delta^*_{k,1}\}}
\cdot \frac{{\cal M}_R\{p\}}{||{\cal M}_R||^2}
= e^{\hat{\cal W}}\cdot 1
\ee
which is naturally graded, and
${\cal Z}_m$ denotes the contribution of the Young diagrams of the size 
(number of boxes) $|R|=m$.

It turns out that from
the point of view of $N$-dependence   the $\hat {\cal W}$ operator
contains just three different structures:
$t^{N_1+N_2}+t^{-N_1-N_2}$, $t^{N_1-N_2}+t^{N_2-N_1}$ and $t^{-N_1-N_2}$:
\be
\boxed{
\begin{array}{c}
\hat {\cal W} = \frac{q^2}{q^2-1}\left\{
\Big(t^{N_1+N_2}+t^{-N_1-N_2}\Big)
\left(\underline{\frac{p_1}{q^2-1}}+\sum_{m=2} \frac{p_m\hat D_{[m-1]}}{q^{2m}-1}\right) +
\right.   \\ \\ \left. +
\Big(t^{N_1-N_2}+t^{N_2-N_1}\Big)
\left(\,\underline{-\frac{p_1}{q^2-1}}+
\sum_{m=2}\frac{p_m\hat D_{m-1}}{q^{2m}-1}
\right)
\right\} \
+ \  \frac{t^{-N_1-N_2}}{q^2-1}\cdot \hat W
\end{array}
}
\label{Ndep}
\ee
with the underlined $p_1$-terms combined into
${\cal Z}_1=\frac{(t^{N_1}-t^{-N_1})(t^{N_2}-t^{-N_2})}{(q-q^{-1})^2}\cdot p_1
=\frac{\{t^{N_1}\}\{t^{N_2}\}}{\{q\}^2}\cdot p_1$
and with a somewhat sophisticated but $N$-independent $\hat W$:
\be
\!\!\!\!\!\!\!\!\!\!\!\!
\begin{array}{c}
\hat W =
\frac{1}{(t^2-1)(q^2-1)}
\sum_{m=2} \frac{mS_{[1^m]}\hat D_{[1^{m-1}]}}{q^{2m-4}}
+  \\ \\
+ \sum_{m=3} \frac{p_m}{q^{2m}-1}\left(
(-2q^2+t^2)\hat D_{[m-1]} + \frac{q^4-t^4}{q^2}\hat D_{[m-2,1]}
+ \frac{t^2(t^2-q^2)}{q^2}\hat D_{[m-3,2]}
+ \frac{t^6-q^6}{q^4}\hat D_{[m-3,1,1]}
+ \right.  \\ \\ \left.
\ \ \ \ \ \ \ \ \ \  +\ \ \ \ldots \ \ \
+ \frac{(-)^m}{q^{2m-6}}\cdot \frac{(t^{2m-2}-1)(q^{2m-2}-1)}{(t^2-1)(q^2-1)}\hat
D_{[1^{m-1}]}
\right)\ \  + \!\!\!\!\!\!\!\!\!\!\!\!\!\! \\ \\
+\sum_{m=3} \frac{p_{m-1}p_1(t^2-1)}{(q^{2m-2}-1) }\left(
- \hat D_{[m-1]} + \frac{t^2 }{q^2}\hat D_{[m-2,1]}
+ \right.  \\ \\ \left.
+ \frac{ (q^2-t^2)}{q^2}\hat D_{[m-3,2]}
- \frac{t^4 }{q^4 }\hat D_{[m-3,1,1]}
+ \ldots
- \frac{(-)^m(t^{2m-4}-1)(q^{2m-4}-1)}{q^{2m-6}(t^2-1)(q^2-1)}\hat D_{[1^{m-1}]}
\right)
+ \ \ldots \ +
\\ \\
+\frac{p_3}{q^6-1}\sum_{m=5} S_{[1^{m-3}]}\cdot
\left( \frac{\hat D_{[m-3,2]}-\hat D_{[m-2,1]}}{q^2}+\frac{t^2\hat D_{[m-3,1,1]}}{q^4}
- \frac{(t^4-1)(q^4-1)}{(t^2-1)(q^2-1)}\cdot \frac{\hat D_{[1^{m-1}]}}{q^{2m-6}}
\right)
+  \\
+\frac{p_2}{q^{4}-1}\sum_{m=4}
\frac{ S_{[1^{m-2}]}\cdot
\big(\hat D_{[1^{m-1}]}-\hat
D_{[2,1^{m-3}]}\big)^{\phantom{\big(\big)}}\!\!\!\!\!\!}{q^{2m-6}}
\!\!\!\!\!\!\!\!\!\!\!\!\!\!\!\!\!\!\!\!\!\!\!\!\!\!\!\!\!\!\!\!\!\!\!\!
\end{array}
\label{firstterms}
\ee
In these formulas we use the condensed notation,
which is natural already at the level of (\ref{HamCR}):
\be
S_R={\rm Schur}_R\left\{(t^{2k}-1)p_k\right\}, \ \ \ \ \ \ \ \
\hat D_k = (q^{2k}-1)k\frac{\p}{\p p_k}
\ \ \ \ \ {\rm and} \ \ \ \ \
\hat D_R = {\rm Schur}_R\{\hat D_k\}
\ee
Note that the first line in (\ref{Ndep}) contains a Schur polynomial
$\hat D_{[m-1]}$, while the second line -- just a single derivative $\hat D_{m-1}$.
It is clear that one more combination actually appears in these formulas,
\be
P_k = \frac{p_k}{q^{2k}-1},   \ \ \ \ {\rm so \ that} \ \ \ \
\hat D_k = k\frac{\p}{\p P_k}
\ee
However, before making use of this observation, one should answer another obvious question.

\section{Ambiguities at generic $q,t$
\label{ambigqt}}

Comparing sophisticated (\ref{firstterms}) with the simply-looking (\ref{Ndep})
it is natural to ask, if one can use the ambiguity in $\hat {\cal W}$ to simplify $\hat W$.
This desire is only strengthened by the fact that
the result (\ref{Ndep}), though very nice from all other perspectives,
makes fighting ambiguity after the full-fledged $q,t$-deformation  more difficult.
Indeed, in sec.\ref{ambigbeta} we fixed the ambiguity by restricting the $N$-dependence,
but now this is impossible -- all ambiguities preserve the fact that there are
just three different structures:
$t^{N_1+N_2}+t^{-N_1-N_2}$, $t^{N_1-N_2}+t^{N_2-N_1}$ and $t^{-N_1-N_2}$.
In fact, and it is the $\hbar$-expansion which converts them into many independently-looking
structures in the $\beta$-deformed formulas.

What saves the situation is that the ambiguities are still in one-to-one correspondence
with those in sec.\ref{ambigbeta}, in particular the first three are

{\footnotesize
\be
\sigma_3\cdot \left\{-q^2\Big(t^{N_1+N_2}+t^{-N_1-N_2}\Big)\hat D_{[1,1]}
 {- q^2\Big(t^{N_1-N_2}+t^{N_2-N_1}\Big)\hat D_{[2]}}
+ \frac{1}{t^{N_1+N_2}}\Big((q^2+t^2)\hat D_{[1,1]} + (q^2t^2+1)\hat D_{[2]}\Big)
\right.
+ \nn \\
+\left.\frac{(t^2-1)p_1}{t^{N_1+N_2}}\Big(q^2\hat D_{[3]}+(q^2t^2+1)\hat D_{[2,1]}
+ t^2\hat D_{[1,1,1]}\Big)
+ \ldots
\right\} \ + \!\!\!\!\!\!\!\!
%\!\!\!\!\!\!\!\!\!\!\!\!\!\!\!\!\!\!\!\!\!\!\!\!\!\!\!\!\!\!\!\!\!\!\!\!\!\!\!
\nn
\ee
}
{\footnotesize
\be
+\sigma_{4a}\left\{(t^{N_1+N_2}+t^{-N_1-N_2})\hat D_{[2,1]}
+  {(t^{N_1-N_2}+t^{N_2-N_1})\hat D_{[3]}}
+ \right.\nn \\ \left.
+\frac{t^{-N_1-N_2}}{q^4}\Big(-q^2(q^2t^2-t^2+1)\hat D_{[3]}
+(q^2t^4-q^4-q^2t^2+t^2)\hat D_{[2,1]} + t^4\hat D_{[1,1,1]} + \ldots\Big)
\right\}
+ \!\!\!\!\!\!\!\!\!\!\!\!\!\!\!\!  \nn \\
+\sigma_{4b} \left\{
(t^{N_1+N_2}+t^{-N_1-N_2})\hat D_{[1,1,1]}
  {-(t^{N_1-N_2}+t^{N_2-N_1}) \hat D_{2,1}}
  - \right.\nn \\ \left.
 - \frac{t^{-N_1-N_2}}{q^4}\Big(-q^2(q^2t^2-q^2+1)\hat D_{[3]}
 - (q^2t^2+1)^2\hat D_{[2,1]} - (q^2t^4+q^4+t^2)\hat D_{[1,1,1]} + \ldots\Big)
\right\}   \!\!\!\!\!\!\!\!\!\!\!\!\!\!\!\!
\nn
\ee
}

\noindent
to be compared with (\ref{ambWJC3}) and (\ref{ambWJC4})
(with $\sigma$'s differing by rescalings and linear combinations).
This means that one can eliminate the ambiguity for arbitrary $q,t$
by simply requiring that all $\sigma=0$,
i.e., with a certain abuse of terminology,
that the $\beta$-deformed counterpart of the $W$-representation is (\ref{betaC}).
Exact specification of our choice (\ref{firstterms}) is that all $\sigma=0$.

\section{Condensed form of the W-representation
}

Coming back to (\ref{firstterms}) one can see that the operator
$\hat W$ has a peculiar tri-linear structure:
\be
\hat W = \sum_{m=2}\sum_{Q\vdash m-1}\sum_k
C_{k}^Q\cdot P_k \cdot S_{[1^{m-k}]} \cdot \hat D_Q
\label{trilin}
\ee
with  relatively simple $q,t$-dependent (but $p$-independent) coefficients $C_k^Q$.
Indeed,
\be
\hat W = \sum_{m=2}\left\{
\Big((t^2-2q^2)P_m - P_{m-1}\cdot S_{[1]}\Big)\cdot\hat D_{[m-1]}
+ \ \ldots \
+\left(\sum_{k=1}^m \frac{\rho_k p_k}{q^{2k}-1}\cdot S_{[1^{m-k}]}\right)
\cdot\hat D_{[1^{m-1}]}
\right\}
+ \nn \\
+ \Big((q^4-t^4)P_m + t^2P_{m-1}S_{[1]}-P_{m-2}S_{[1,1]}\Big)
\cdot\frac{\hat D_{[m-2,1]}}{q^2}
+\nn\\
+ \Big(t^2(t^2-q^2)P_m - (t^2-q^2)P_{m-1}S_{[1]}+P_{m-2}S_{[1,1]}\Big)
\cdot\frac{\hat D_{[m-3,2]}}{q^2}
+ \nn\\
+ \Big((t^6-q^6)P_m - t^4P_{m-1}S_{[1]}+t^2P_{m-2}S_{[1,1]}-P_{m-3}S_{[1,1,1]}\Big)
\cdot\frac{\hat D_{[m-2,1]}}{q^4}
+ \ldots
\ee
where
\be
\rho_k=
(-)^{k+1}\cdot\left(t^{2k-2}\cdot\frac{q^{2k}-1}{q^2-1}+\frac{t^{2k-2}-1}{t^2-1}\right)
= (-)^{k+1}\cdot\left(q^2t^{2k-2}\cdot\frac{q^{2k-2}-1}{q^2-1}+\frac{t^{2k}-1}{t^2-1}\right)
\ee

This structure is seen even better, if one uses a further condensed notation:
\be
{\cal P}_m^{(k)} := \sum_{l=0}^k (-)^lt^{2(k-l)}P_{m-l}S_{[1^l]}
\ee
In these terms we can list many more items:
\be
\hat W = \underline{
\sum_{m=2} \left(-q^2{\cal P}_m^{(0)}\cdot\hat D_{[m-1]}
+ \sum_{k=1}^m  \rho_k\, P_k \,S_{[1^{m-k}]}
\cdot\frac{\hat D_{[1^{m-1}]}}{q^{2(m-2)}}\right)
}
+\sum_{m=3} \Big({\cal P}_m^1-
q^2{\cal P}_m^{(0)}\Big)\cdot\hat D_{[m-1]}
-\nn \\
- \sum_{m=4} \Big({\cal P}_m^2-q^4{\cal P}_m^{(0)}\Big)\cdot\frac{\hat D_{[m-2,1]}}{q^2}
+ \sum_{m=5}\left\{ \Big({\cal P}_m^{(2)}-q^2{\cal P}_m^{(1)}\Big)
\cdot\frac{\hat D_{[m-3,2]}}{q^2}
+ \Big({\cal P}_m^{(3)}-q^6{\cal P}_m^{(0)}\Big)\cdot\frac{\hat D_{[m-3,1,1]}}{q^4}
\right\}
+\nn \\
+ \sum_{m=6}\left\{- \Big({\cal P}_m^{(3)}-q^4{\cal P}_m^{(1)}\Big)
\cdot\frac{\hat D_{[m-4,2,1]}}{q^4}
- \Big({\cal P}_m^{(4)}-q^8{\cal P}_m^{(0)}\Big)\cdot\frac{\hat D_{[m-4,1,1,1]}}{q^6}
\right\}
+\nn
\ee
{\footnotesize
\be
\!\!\!\!\!\!\!\!\!\!\!\!\!
+ \sum_{m=7}\left\{-\Big(\overbrace{{\cal P}_m^{(2)}-  {\cal P}_m^{(2)}}^{0}\Big)
\cdot\frac{\hat D_{[m-4,3]}}{q^2}
+  \Big({\cal P}_m^{(3)}- q^{2 }{\cal P}_m^{(2)}\Big)
\cdot\frac{\hat D_{[m-5,2,2]}}{q^4}
+ \Big({\cal P}_m^{(4)}-q^{6}{\cal P}_m^{(1)} \Big)
\cdot\frac{\hat D_{[m-5,2,1,1]}}{q^6}
+ \Big({\cal P}_m^{(5)}-q^{10}{\cal P}_m^{(0)}\Big)\cdot\frac{\hat D_{[m-5,1,1,1,1]}}{q^8}
\right\}
+ \nn \\
\!\!\!\!\!\!\!\!\!\!\!\!\!\!\!\!\!\!\!\!
+ \sum_{m=8}\left\{\Big(\overbrace{{\cal P}_m^{(3)}
-  {\cal P}_m^{(3)}}^{0}\Big)
\cdot\frac{\hat D_{[m-5,3,1]}}{q^4}
-  \Big({\cal P}_m^{(4)}- q^{4 }{\cal P}_m^{(2)}\Big)
\cdot\frac{\hat D_{[m-6,2,2,1]}}{q^6}
- \Big({\cal P}_m^{(5)}-q^{8}{\cal P}_m^{(1)} \Big)
\cdot\frac{\hat D_{[m-6,2,1,1,1]}}{q^8}
- \Big({\cal P}_m^{(6)}-q^{12}{\cal P}_m^{(0)}\Big)\cdot\frac{\hat D_{[m-6,1,1,1,1]}}{q^{10}}
\right\}
+ \nn \\
\!\!\!\!\!\!\!\!
+ \sum_{m=9}\left\{\Big(\overbrace{{\cal P}_m^{(2)}-  {\cal P}_m^{(2)}}^{0}\Big)
\cdot\frac{\hat D_{[m-5,4]}}{q^2}
-\Big(\overbrace{{\cal P}_m^{(3)}-  {\cal P}_m^{(3)}}^{0}\Big)
\cdot\frac{\hat D_{[m-6,3,2]}}{q^4}
-\Big(\overbrace{{\cal P}_m^{(4)}-  {\cal P}_m^{(4)}}^{0}\Big)
\cdot\frac{\hat D_{[m-6,3,1,1]}}{q^6}
+  \Big({\cal P}_m^{(4)}- q^{2 }{\cal P}_m^{(3)}\Big)
\cdot\frac{\hat D_{[m-7,2,2,2]}}{q^6}
+ \ \ \ \ \ \ \ \ \ \ \right.\nn \\ \left.
+ \Big({\cal P}_m^{(5)}-q^{6}{\cal P}_m^{(2)} \Big)
\cdot\frac{\hat D_{[m-7,2,2,1,1]}}{q^8}
+ \Big({\cal P}_m^{(6)}-q^{10}{\cal P}_m^{(1)}\Big)\cdot\frac{\hat
D_{[m-7,2,1,1,1,1]}}{q^{10}}
+ \Big({\cal P}_m^{(7)}-q^{14}{\cal P}_m^{(0)}\Big)\cdot\frac{\hat D_{[m-7, 1^6]}}{q^{12}}
\phantom{\overbrace{{\cal P}_m^{(4)}-  {\cal P}_m^{(4)}}^{0}}
\!\!\!\!\!\!\!\!\!\!\!\!\!\!\!\!\!\!\!\!\!\!\!\!\!\!\!\!\!
\right\}
+ \ldots  \ \ \ \ \ \ \ \ \ \
\nn
\ee
}

\noindent
and it gets clear, that contributing are only derivatives $D_Q$ with Young diagrams
of the type $Q=[s,2^{n_2},1^{n_1}]$, i.e. with at most one line of the length $s>2$
-- exactly those which define the {\it simple} Hurwitz numbers,
see \cite{Ok,MMN1,compmod} and references therein. The size of the diagram is
fixed to be $|Q|=s+2n_2+n_1=m-1$, i.e.
%we can conjecture that
\be
\!\!\!\! \!\!\!\!\!
\boxed{
\hat W =
\underline{
\sum_{m=2} \left(-q^2{\cal P}_m^{(0)}\cdot\hat D_{[m-1]}
+ \sum_{k=1}^m   \rho_k\, P_k \,S_{[1^{m-k}]}
\cdot\frac{\hat D_{[1^{m-1}]}}{q^{2(m-2)}}\right)
}
+ \!\!\! \!\!\! \!\!\!\!\sum_{\stackrel{s\geq 2}{\stackrel{n_1,n_2\geq 0}{s+2n_2+n_1=m-1}}}
 \!\!\! \!\!\!\!\!\!\!\!\!
(-)^{n_1}\Big({\cal P}_{m}^{(n_1+n_2+1)}-q^{2(n_1+1)}{\cal P}_m^{(n_2)}\Big)
\frac{\hat D_{[s,2^{^{n_2}},1^{^{n_1}}]}}{q^{2(n_1+n_2)}}
%+\!\!\!\sum_{\stackrel{m=2}{\stackrel{Q\vdash m-1}{\underline{Q_1>1}}}}\!\!
%(-)^{m-Q_1-1}\cdot
%\Big( {\cal P}_m^{(l_Q)} -q^{2(l_Q-\nu_Q)} {\cal P}_m^{(\nu_Q)}\Big)
%\cdot \frac{\hat D_Q}{q^{2l_Q-2}}
} \
\label{sumP}
\label{genfun}
\ee
Underlined are the items, which deviate from the general rule and
need a separate treatment.
We can attempt to improve the last formula by lifting restriction $s\geq 2$
on the length of the first row in the diagram $Q$ (when $n_2=0$).
This requires switching from $\rho_k$ to a
slightly nicer $\tilde\rho_k$, with the help of the following manipulation:
\be
\left.\begin{array}{c}
\sum_{k=1}^m \rho_k\, P_k \,S_{[1^{m-k}]}
= \sum_{k=1}^m (-)^{k+1}
\overbrace{\left(t^{2k-2}\cdot\frac{q^{2k}-1}{q^2-1}+\frac{t^{2k-2}-1}{t^2-1}\right)}^{\rho_k}
 P_k \,S_{[1^{m-k}]} \\ \\
{\cal P}_m^{(m-1)} = \sum_{l=0}^{m-1} (-)^l t^{2(m-1-l)}P_{m-l}S_{[1^l]}
\ \stackrel{l=m-k}{=}\  \sum_{k=1}^m (-)^{m-k} t^{2k-2}\,P_{k}\,S_{[1^{m-k}]}
\end{array}\right\} \ \Longrightarrow
\nn \\
\sum_{k=1}^m \rho_k\, P_k \,S_{[1^{m-k}]}
= (-)^{m}
{\cal P}_m^{(m-1)}
+ \sum_{k=1}^m (-)^{k+1}
\underbrace{\left(t^{2k-2}\cdot\frac{q^{2k}-1}{q^2-1}+\frac{t^{2k}-1}{t^2-1}
\right)}_{\tilde\rho_k}
 P_k \,S_{[1^{m-k}]}
\ee
However, this is not sufficient to provide the second term
in the would-be coefficient ${\cal }P_m^{(m-1)}\!-q^{2m}{\cal P}_m^{(0)}$ of $\hat
D_{[1^{m-1}]}$.

\section{From derivatives to shifts}

So far we expanded $\hat W$ operators in $p$-derivatives.
However, after $q$-deformation one can instead consider difference operators
(shifts of $p_k$).
Transformation between derivatives to shifts is provided by the Cauchy formula
\cite{Macdonald}, recently reviewed from related perspective in \cite{Cauchy},
which expresses shifts through Schur polynomials of derivatives:
\be
\exp\left(\sum_k  \bar p_k \frac{\p}{\p p_k}\right)
= \sum_R {\rm Schur}_R\{\bar p_k\} \cdot {\rm Schur}_R\left\{k\frac{\p}{\p p_k}\right\}
\ee
in particular, for $\bar p_k=z^k$, we get a contribution from only single-line Young diagrams
$R=[m]$ (symmetric representations):
\be
\exp\left(\sum_k  z^k \frac{\p}{\p p_k}\right)
= \sum_m z^m \cdot {\rm Schur}_{[m]}\left\{k\frac{\p}{\p p_k}\right\}
\ee
Changing sign, we can restrict to single-row diagrams $R=[1^m]$ (antisymmetric
representations):
\be
\exp\left(-\sum_k  z^k \frac{\p}{\p p_k}\right)
= \sum_m (-)^mz^m \cdot {\rm Schur}_{[1^m]}\left\{k\frac{\p}{\p p_k}\right\}
\ee
Tri-linear structure (\ref{trilin}) implies that in addition to two exponentials
in (\ref{HamCR}), providing Schur functions of $(t^{2k}-1)p_k$ and $\hat D_k$
one needs an additional insertion like $\sum_k P_kz^{-k} = \sum_k\frac{p_kz^{-k}}{q^{2k}-1}$
in the integrand.
However, more important is that
in (\ref{sumP}) contributing are diagrams $Q$ with arbitrary number of rows $l_Q$,
thus the single-$z$ integrals like (\ref{HamCR}) should be substituted by
multiple-$z$ integrals
(because of the restriction on the line lenghts, one can actually hope to
manage with finitely many $z$-variables).
Amusingly, such integrals are also expected in description of generalized
Kerov functions \cite{Kerov} -- and the corresponding theory still remains to be built.

\section{Conclusion}

In this paper we conjectured a prototype (\ref{genfun})
of the $W$-representation of the  basic $q,t$-model,
generalizing the Gaussian  complex-matrix models.
As suggested in \cite{MPSh} we used as a definition of $q,t$-model
a sum  of Macdonald averages, defined through the character-preservation postulate
of \cite{MMchar}.
Exact relation to network formulation of \cite{networkZ} remains to be worked out,
together with a possible derivation from the $W$-representation
of the deep ${\cal R}$-matrix structures and Knizhnik-Zamolodchikov equations,
which so far are revealed (also, just partly)
only in the network-model approach \cite{qtKZ}.
An important step in this direction can be further lifting to
generalized Macdonald functions, depending on collections of Young diagrams.
which are capable to describe many free fields and thus the full-fledged
networks with many horizontal lines.

Our resulting formula (\ref{genfun}) is hardly final -- rather an intermediate one.
It is still far from ideal
and should be further improved and converted into something, probably more similar
to the Calogero-Ruijsenaars Hamiltonian (\ref{HamCR}).
In continuous models
$W$-representations always involve just another harmonic of the same operator,
whose zero-harmonic provides the Hamiltonian -- of which the relevant characters
were the eigenfunctions.
Surprisingly or not, a similar relation in the $q,t$-deformed case is far from obvious.
This should have something to do with the fact that the relevant integrals
in network models are multiple Jackson integrals,
which are actually sums over (collections of) Young diagrams --
and the notion of harmonics in this case is not fully obvious.
Further improvement (condensed notation) of (\ref{genfun}) should be an important,
and, perhaps, the most straightforward step towards building appropriate formalism.
An important fact in this relations is that the asymmetry between $q$ and $t$,
which is very strong in the starting definition (\ref{charavMC}) of the $q,t$ model,
almost disappears at the level of the $N$-independent $W$-operator in (\ref{genfun}) --
as one expects in the context of network models.
Of separate interest and significance is the detailed study of particular limits --
not only to $\beta$-deformed ($t=q^\beta\longrightarrow 1$), but also to
the Hall-Littlewood ($t=0$) case.

\section*{Acknowledgements}

My work is partly supported by the grant of the
Foundation for the Advancement of Theoretical Physics BASIS,
by RFBR grant 19-02-00815
and by the joint grants 17-51-50051-YaF, 18-51-05015-Arm,
18-51-45010-Ind,  RFBR-GFEN 19-51-53014.
I also acknowledge the hospitality of KITP and
partial support by the National Science Foundation
under Grant No. NSF PHY-1748958
at an early stage of this project.

\end{document}